\documentclass[twocolumn,floatfix,altaffilletter,preprintnumbers,
               tightenlines,showpacs,showkeys,nofootinbib]{revtex4-1}
\usepackage[utf8]{inputenc}
\usepackage[colorlinks=true,citecolor=blue,linkcolor=blue]{hyperref}
\usepackage[normalem]{ulem}
\usepackage{amsmath,amssymb}
\usepackage{epsfig}
\usepackage{graphicx}               
\usepackage{url}
\usepackage{color}
\usepackage{slashed}
\usepackage{multirow}
\usepackage{placeins}
\usepackage[dvipsnames]{xcolor}
\usepackage{epstopdf}
\usepackage{soul}
\usepackage{tikz}
\usepackage[capitalise, english]{cleveref}
\usepackage{siunitx}
\usepackage{xspace}
\usetikzlibrary{trees}
\usetikzlibrary{decorations.pathmorphing}
\usetikzlibrary{decorations.markings}

\newcommand\myshade{80}
\colorlet{mylinkcolor}{ForestGreen}
\colorlet{mycitecolor}{Red}
\colorlet{myurlcolor}{violet}

\hypersetup{
  linkcolor  = mylinkcolor!\myshade!black,
  citecolor  = mycitecolor!\myshade!black,
  urlcolor   = myurlcolor!\myshade!black,
  colorlinks = true
}

\definecolor{jblue}{RGB}{20,50,100}
\definecolor{npurple}{RGB} {153, 51, 204}
\definecolor{wred}{RGB}{217,0,56}
\definecolor{white}{RGB}{255,255,255}

\definecolor{korange}{RGB}{235, 80,  43}
\definecolor{korange2}{RGB}{245, 100,  63}
\definecolor{kyelloworange}{RGB}{255, 210,  110}
\definecolor{kyelloworange2}{RGB}{240, 170,  90}
\definecolor{kred}{RGB}{204,  102, 153}
\definecolor{kpurple}{RGB}{153,  61, 190}
\definecolor{kpurplelight}{RGB}{213,  161, 230}

\DeclareSIUnit\year{yr}
\DeclareSIUnit\pc{pc}
\DeclareSIUnit\ergs{ergs}
\DeclareSIUnit\msun{\ensuremath{M_\odot}}
\allowdisplaybreaks

\newcommand{\eps}{\epsilon}

\def\thesection{\Roman{section}}

\newcommand{\AddrMainz}{%
PRISMA$^{+}$ Cluster of Excellence \& Mainz Institute for Theoretical Physics,\\
Johannes Gutenberg-Universit\"at Mainz, 55099 Mainz, Germany
}

\begin{document}
\title{Gravitational wave probes of axion-like particles}

\author{Camila S. Machado}
\email{camachad@uni-mainz.de}
\author{Wolfram Ratzinger}
\email{w.ratzinger@uni-mainz.de}
\author{Pedro Schwaller}
\email{pedro.schwaller@uni-mainz.de}
\author{Ben A. Stefanek}
\email{bstefan@uni-mainz.de}
\affiliation{\normalsize\it \AddrMainz}

\preprint{MITP/19-083 }
\begin{abstract}

We have recently shown that axions and axion-like particles (ALPs) may emit an observable stochastic gravitational wave (GW) background when they begin to oscillate in the early universe. In this note, we identify the regions of ALP parameter space which may be probed by future GW detectors, including ground- and space-based interferometers and pulsar timing arrays. Interestingly, these experiments have the ability to probe axions from the bottom up, i.e. in the very weakly coupled regime which is otherwise unconstrained. Furthermore, we discuss the effects of finite dark photon mass and kinetic mixing on the mechanism, as well as the (in)sensitivity to couplings of the axion to Standard Model fields. We conclude that realistic axion and ALP scenarios may indeed be probed by GW experiments in the future, and provide signal templates for further studies.

\end{abstract}

\maketitle

\section{Introduction}
The direct detection of gravitational waves (GW) opened up a new avenue to explore fundamental physics in the early universe. In particular, axions or axion-like particles (ALPs) are a well-motivated extension of the Standard Model (SM), e.g. to solve the strong CP problem~\cite{PhysRevLett.38.1440}, to provide a dynamical solution to the electroweak hierarchy problem~\cite{Graham:2015cka}, to provide suitable inflaton~\cite{PhysRevLett.65.3233} or dark matter (DM) candidates~\cite{Preskill:1982cy,Abbott:1982af,Dine:1982ah}, or in the context of string theory~\cite{Arvanitaki:2009fg}.  Experimental searches for these particles are covering an increasing part of the parameter space. Several searches rely on the axion-photon coupling which is generically inversely proportional to the axion decay constant. This means the region corresponding to smaller decay constants (larger couplings) is more constrained, whereas larger values are usually difficult to probe. 

In Ref.~\cite{Machado:2018nqk}, we showed that axions or axion-like particles coupled to a light dark photon can produce a stochastic gravitational wave background (SGWB) when the axion field begins to oscillate in the early universe, allowing exploration of parameter space inaccessible to experiments that rely on the axion-photon coupling. The rolling axion induces a tachyonic instability that amplifies vacuum fluctuations of a single gauge boson helicity, sourcing chiral GWs. The energy transfer from the axion into light vectors also widens the viable parameter space for axion DM. 

The goal of this paper is to explore the phenomenological impact of our findings. First, we show that GWs can be produced in realistic axion and ALP scenarios where other couplings such as kinetic mixing of the SM and dark photon, couplings of the axion to SM fields, or non-zero dark photon masses are also present. Next, we provide a simple analytic fit to the GW spectrum extracted from our numerical simulation, useful for further studies or comparison with GW signals from other sources. We present the main result of our paper in Fig.~\ref{fig:wolframs_money}, where we identify the regions of ALP parameter space that will be probed by future GW experiments. Since a strong polarization of the GW signal peak is a firm prediction of our scenario, in \cref{fig:bounds} we indicate the region where this feature may be probed following the recent results of Ref.~\cite{Domcke:2019zls}.  It is striking that gravitational waves may be able to provide evidence for axions with very large decay constants which are otherwise inaccessible.

\section{The Audible Axion Model}
Here we give a brief overview of the model presented in Ref.~\cite{Machado:2018nqk}, which consisted of an axion field $\phi$ and a massless dark photon $X_{\mu}$ of an unbroken $U(1)_X$ Abelian gauge group
\begin{equation}
\begin{split}
\frac{\mathcal{L}}{\sqrt{-g}} =\frac{1}{2}\partial_{\mu} \phi \,\partial^{\mu} \phi - V(\phi) -\frac{1}{4} X_{\mu\nu} X^{\mu\nu} - \frac{\alpha}{4f}\phi X_{\mu\nu}\widetilde{X}^{\mu\nu}  \, ,
\end{split}
\label{lag}
\end{equation}
where the parameter $f$ is the scale at which the global symmetry corresponding to the Nambu-Goldstone field $\phi$ is broken~\footnote{We consider $\alpha>1$ in order to have efficient particle production, which can be obtained in several UV completions, see e.g. \cite{Agrawal:2017cmd}.}. We assume this global symmetry is also explicitly broken at the scale $\Lambda \sim \sqrt{m\,f}$, resulting in the potential $V(\phi)$ and a mass $m$ for the axion. 

While the expansion rate of the universe $H = a'/a^{2}$ is greater than the axion mass $m$, the axion field is overdamped and does not roll~\footnote{Here, primes denote derivatives with respect to conformal time.}. In a radiation-dominated universe, $H$ becomes of order $m$ at the temperature $T_{\rm osc} \approx \sqrt{m M_{P}}$, at which point the axion will begin to oscillate in its potential with initial conditions given by misalignment arguments, namely $\phi_i= \theta f$, $\phi_i^{\prime} \approx 0$, and $\theta \sim \mathcal{O}(1)$,
where $\theta$ is the initial misalignment angle.  The $\phi X_{\mu\nu}\widetilde{X}^{\mu\nu}$ coupling results in a non-trivial dispersion relation for the gauge field helicities
\begin{equation}
\omega^{2}_{\pm}(k, \tau)= k^{2}  \mp k\frac{\alpha}{f} \phi'\,,
\end{equation}
that depends explicitly on the velocity $\phi'$ of the axion field. As the axion field oscillates, one of the gauge field helicities will have a range of modes with imaginary frequencies (negative $\omega^2$), resulting in a tachyonic instability that drives exponential growth. This process transfers energy from the axion field into dark gauge bosons and amplifies vacuum fluctuations of the tachyonic modes into a rapidly time-varying, anisotropic energy distribution that sources GWs. 
For an in-depth review of the particle production process and its applications, see Refs.~\cite{Ratra:1991bn,Garretson:1992vt,Field:1998hi,Lee:2001hj,Campanelli:2005ye,Anber:2006xt,Anber:2009ua,Barnaby:2010vf,Barnaby:2011vw,Barnaby:2011qe,Barnaby:2012tk,Adshead:2013qp,Adshead:2015pva,Giblin:2017wlo,Hook:2016mqo,Domcke:2016bkh,Kitajima:2017peg,Agrawal:2017eqm,Fonseca:2018xzp,Dror:2018pdh,Co:2018lka,Bastero-Gil:2018uel,Agrawal:2018vin,Soda:2017dsu,Kitajima:2018zco,Carenza:2019vzg,Alonso-Alvarez:2019ssa,Adshead:2018doq,Adshead:2019igv,Adshead:2019lbr}. We will now briefly discuss some possible extensions to the original simplified model.

\subsection{Finite dark photon mass}
\label{sec:mXneq0}
First, we consider the possibility of a non-zero mass for the dark photon
 which could arise through a dark Higgs or Stueckelberg mechanism. The main effect of $m_{X}$ is to modify the dark photon dispersion relation
 \begin{equation}
\omega^{2}_{\pm}(k, \tau)= k^{2}  + a^{2} m_{X}^{2} \mp k\frac{\alpha}{f} \phi'\,,
\label{eq:dispersion}
\end{equation}
which can reduce the efficiency of or prevent tachyonic growth.
To further quantify this statement, we go back to the analysis in Ref.~\cite{Machado:2018nqk}, where we showed that the tachyonic growth of the mode functions becomes inefficient if they grow less than $\mathcal{O}(1)$ during one oscillation of the axion field. This happens when $-\omega^2_{\pm}<(am)^2$ is satisfied for all modes $k$. From this we can deduce that for $\alpha\theta \gtrsim 10$, we require $m_{X} \lesssim \theta\alpha m / 2$ in order to have tachyonic production. Here, we focus on dark photon masses well below this bound which will not affect the success of our mechanism. The case where this is not true is discussed in~\cref{app:finiteMass}.

\subsection{Kinetic Mixing}
Next, we examine whether the relevant photon-dark photon kinetic mixing operator
\begin{equation}
\Delta\mathcal{L} =  - \frac{\eps}{2} F_{\mu\nu} X^{\mu\nu} \,,
\label{eq:kinMix}
\end{equation}
affects our mechanism. Indeed, this operator will inevitably be generated by renormalization group flow if there exist states which carry both electromagnetic and $U(1)_X$ charge~\cite{Holdom:1985ag}. If kinetic mixing leads to an effective coupling of the dark photon to the SM radiation bath, one might worry that it
induces a large thermal mass for the dark photon that prevents tachyonic growth. 

In the case of an exactly massless dark photon $m_X=0$, the kinetic mixing term is unphysical as it can be removed via the field redefinition $X'=X+\epsilon A$ and $A'=A/\sqrt{1-\epsilon^2}$
that leaves the coupling of the SM photon to the electromagnetic current unchanged. Thus, it is clear that only the field combination that couples to the SM plasma $A'$ develops a thermal mass.

However, for $m_X\neq 0$, the mixing is physical. Diagonalizing the kinetic terms by performing the same field redefinition now leads to a non-diagonal mass matrix which, in addition to the thermal mass $\Pi$ induced by the SM plasma for $A'$, must be included in the dispersion relation
\begin{equation}
 \bigg[\omega^2+k^2+\begin{pmatrix}
 \epsilon'^{2} m_X^2 + \Pi & -\epsilon' m_X^2 \\
 -\epsilon' m_X^2  & m_X^2
\end{pmatrix}\bigg]
\begin{pmatrix} A^{\prime\mu} \\ X^{\prime\mu} \end{pmatrix}=0 \,,
\label{eq:EMM}
\end{equation}
with $\epsilon'=\epsilon/\sqrt{1-\epsilon^2}$. The photon thermal mass is of order $\Pi\approx e^2T^2$, which at the time when the axion begins to oscillate evaluates to $\Pi\approx e^2 m M_P$. As discussed in \cref{sec:mXneq0}, the existence of the tachyonic instability requires $m_X  \lesssim \theta\alpha m / 2$. Futhermore, the momenta that experience tachyonic growth are those with $k\lesssim\theta\alpha m$, so we are deeply in the regime where $ m_{X}^{2}\,, k^{2} \ll \Pi$. In this limit, the effective mass matrix in \cref{eq:EMM} always has a small eigenvalue $m_X^2(1+\mathcal{O}(\epsilon^2))$ which is independent of $T^2$, despite the kinetic mixing~\cite{Dubovsky:2015cca}~\footnote{This result, while perhaps surprising at first, becomes clear when we consider the limit $m_{X} \rightarrow 0$, where the dark photon must decouple.}. Thus, we conclude that the field combination associated with the dark photon $X'$ does not acquire a thermal mass via kinetic mixing, so we are subject only to the usual constraints on $\epsilon$, see e.g. Refs.~\cite{Jaeckel:2010ni,Jaeckel:2007ch,Abel:2006qt,Abel:2008ai,Ahlers:2008qc,Redondo:2008en,Redondo:2008aa,Mirizzi:2009iz,Essig:2009nc,Batell:2009di,Afanasev:2008fv,Jaeckel:2008sz,Gninenko:2008pz,Baryakhtar:2018doz}.

\subsection{QCD Axion}
Finally, we examine the case where the ALP $\phi$ is taken to be the QCD axion itself, which is the focus of Ref.~\cite{Agrawal:2017eqm}. In this limit, $m$ and $f$ are not independent parameters but are instead related by $m^{2} f^{2}= \chi_{\rm QCD}$, where $\chi_{\rm QCD} = (75.5 \,\, {\rm MeV})^{4}$ is the QCD topological susceptibility. In particular, the QCD axion has the following couplings to SM gauge bosons
\begin{equation}
\Delta\mathcal{L} = \frac{\alpha_s }{8\pi f} \phi \, G_{\mu\nu}^a\tilde{G}^{a\,\mu\nu}+\frac{g_{\phi\gamma\gamma}}{4}\phi F^{\mu\nu}\tilde{F}_{\mu\nu} \,,
\end{equation}
where $ G_{\mu\nu}^a$ and $F_{\mu\nu}$ are the gluon and photon field strengths, respectively, and $g_{\phi\gamma\gamma}$ is a model dependent coupling, e.g. $g_{\phi\gamma\gamma} = -1.92\,\alpha_{\rm EM} / (2\pi f)$ in the KSVZ model~\cite{PhysRevLett.43.103,SHIFMAN1980493}.  Here, we note that none of these couplings spoil the effectiveness of our mechanism because the tachyonic growth of these states are regulated by plasma effects. The photon acquires a Debye mass of order $\Pi \sim e^{2} T^{2}$ via hard thermal loops, preventing tachyonic growth~\cite{Kapusta:2006pm,Kraemmer:1995qe}. Similarly, the gluon self-coupling induces a magnetic mass $m(T) \sim g^{2}T$~\cite{LINDE1980289,RevModPhys.53.43,Espinosa:1992kf}. As a final consideration, model dependent couplings of $\phi$ to SM fermions also exist. However, the production of fermions is not exponential due to Pauli-blocking.  Thus, the exponential production of dark photons dominates over SM channels.

\section{Gravitational Wave Spectrum}
\label{sec:GW}

Here, we present an improved computation of the GW spectrum as compared to the results of Ref.~\cite{Machado:2018nqk}. The computation requires the discretization of a double integral over the tachyonic momenta, resulting in the simulation time growing as the square of the number of gauge modes. By switching to a more memory efficient code written in Python, we were able to solve the coupled axion-gauge boson equations of motion using $N=10^{5}$ gauge modes. For the GW spectrum computation, an $N_{\rm GW} =500$ subset of these modes were taken, which is an order of magnitude improvement over our previous simulation.
\begin{figure}[h]
\centering
\includegraphics[width=0.90\columnwidth]{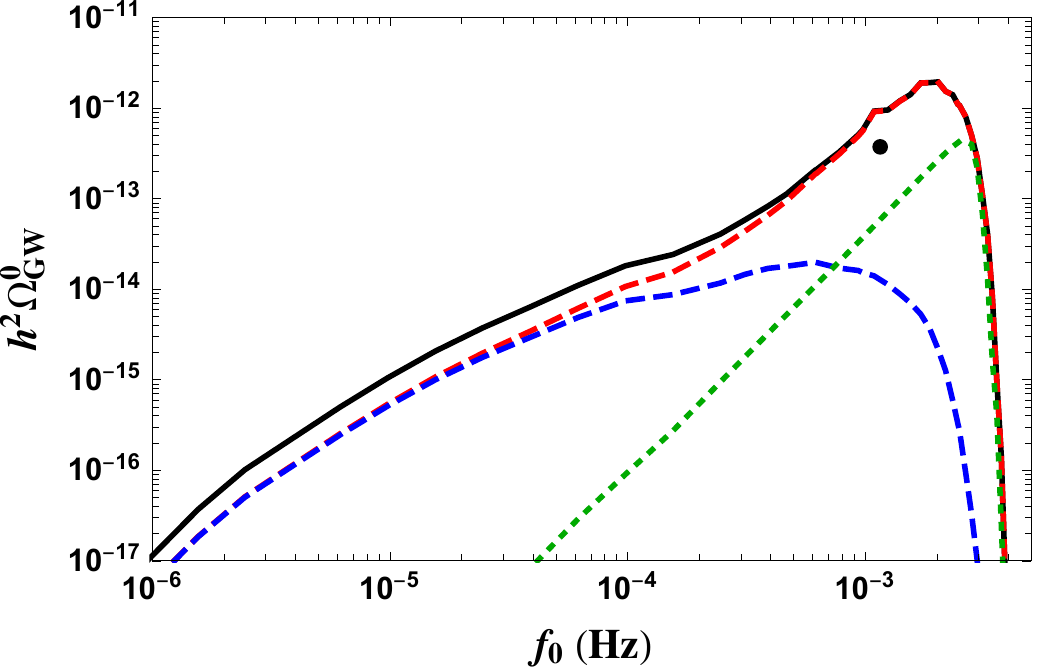}
\caption{GW spectra from the improved numerical simulation for the ALP2 benchmark point of Ref.~\cite{Machado:2018nqk}. Black is the total spectrum, whereas red and blue are the individual polarizations. Green gives the ``conservative spectrum" as described in \cref{sec:GW}.}
\label{fig:GWspec}
\end{figure}
We show the results of the improved numerical calculation in \cref{fig:GWspec}. 
The spectrum is strongly polarized in the peak region, whereas the tail is unpolarized as shown in the figure. Note that backscattering effects (not included here) could affect the degree of polarization at high frequencies~\cite{Adshead:2018doq}. Furthermore, the dashed green line indicates a ``conservative spectrum", which is the part we expect to remain even if the process of gauge bosons backscattering into axions results in a strong back-reaction~\footnote{This effect, which we neglect here, induces inhomogeneities in the axion field which can end the energy transfer from effects which rely on coherent resonance of gauge modes with the zero-momentum axion condensate.}. In any case, we expect the GWs produced during the initial tachyonic instability phase to survive, and we use the estimate for the closure of the tachyonic band given in Ref.~\cite{Machado:2018nqk} to obtain the conservative spectrum shown in \cref{fig:GWspec}. 
\begin{figure*}[t!]
\centering
\includegraphics[width=0.8\textwidth]{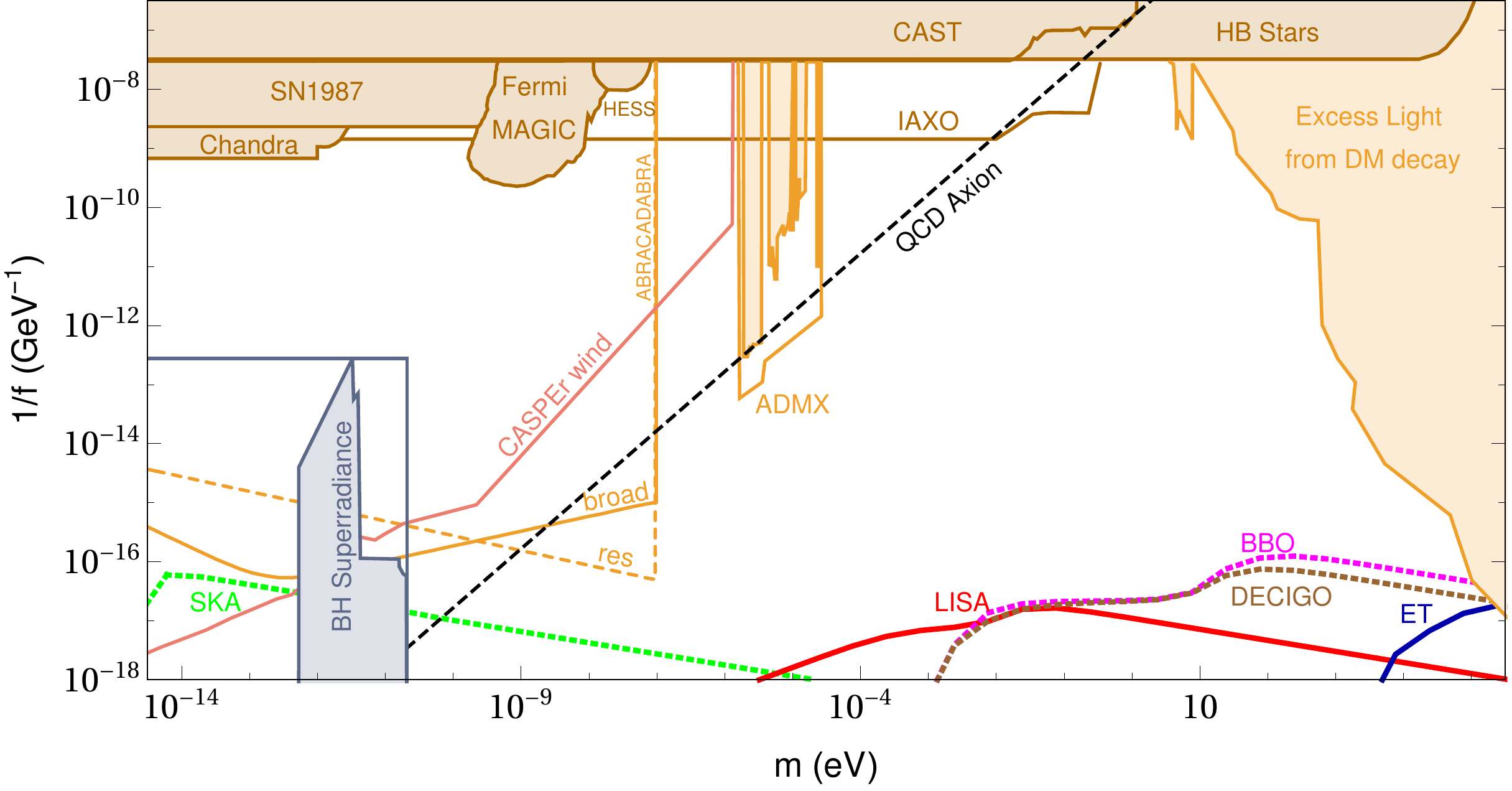}
\caption{Axion and ALP parameter space in the mass vs. inverse decay constant plane. Regions below the colored curves are in reach of future ground-based (ET) and satellite-based (LISA, BBO, DECIGO) GW detectors, or future pulsar timing arrays (SKA). Shaded regions are excluded by existing constraints, while unshaded regions show the sensitivity of various other planned experiments. Black hole superradiance excludes the grey shaded region, and future black hole observations could extend this region to the grey line. The location of the QCD axion band is indicated by the black dashed line.}
\label{fig:wolframs_money}
\end{figure*}
\subsection{GW Spectrum Fit Template}
\label{sec:fit_temp}
To make connection with experimental searches for SGWBs, we use the improved numerical simulation to extract a GW signal template. Such a template enables simple estimates of the GW spectrum and signal-to-noise (SNR) calculations for a given set of model parameters, without having to run a complicated numerical simulation. We approach our GW signal template from the ansatz that the low frequency part of the GW spectrum is given by a power law while the high frequency part falls off exponentially, with some transition region that gives the peak. A reasonable ansatz of this form is
\begin{align}
	\tilde{\Omega}_{\rm GW}(\tilde{f}) = \frac{\mathcal{A}_{s}\left(\tilde{f}/f_{s}\right)^p}{1+\left(\tilde{f}/f_{s}\right)^{p} \exp\left[\gamma (\tilde{f}/f_{s}-1)\right]}\,,
	\label{eq:temp}
\end{align}
where $\tilde{\Omega}_{\rm GW} \equiv \Omega_{\rm GW}(f)/ \Omega_{\rm GW}(f_{\rm peak})$ and $\tilde{f} \equiv f/f_{\rm peak}$. In Ref.~\cite{Machado:2018nqk}, we derived simple analytic scaling relations for the peak amplitude and frequency of the GW spectrum, which at the time of GW emission are
\begin{equation}
f_{\rm peak} \approx (\alpha\theta)^{2/3} m \,, \hspace{2mm} \, \Omega_{\rm GW}(f_{\rm peak})  \approx \left( \frac{f}{M_P} \right)^4 \, \left( \frac{\theta^{2}}{\alpha}\right)^{\frac{4}{3}}  \,,
\label{eq:SRest}
\end{equation}
where these expressions hold for $\alpha \sim 10-100$.
The parameters $\mathcal{A}_{s}$ and $f_{s}$ are fit to the GW spectrum from our numerical simulation to correct for the $\mathcal{O}(1)$ factors by which the scaling relation is off. The parameter $p$ specifies the power law index and $\gamma$ controls how quickly the exponential behavior takes over at high frequencies. Discussion of the fit to the simulation and the best fit values for the parameters $A_{s}, f_{s}, \gamma, p$ can be found in~\cref{app:GWfit}. Together, \cref{eq:temp,eq:SRest} allow one to go directly from the underlying fundamental model parameters $\alpha, m, f$ to the GW spectrum.

\section{Probing Audible Axion Models}
\label{sec:probeAA}
With the results of the previous section, we can now identify the regions of parameter space that may be probed by future GW experiments. Detectability requires an SNR above a certain experiment dependent threshold. Here, we use the values and method of Ref.~\cite{Breitbach:2018ddu}. 
Our results are shown in \cref{fig:wolframs_money}, where the detectable regions lie below the curves labeled as SKA, LISA, BBO, DECIGO and ET, respectively~\footnote{For experiments which probe the axion-photon coupling $g_{\phi\gamma\gamma}$, we assume the KSVZ relation $g_{\phi\gamma\gamma} = -1.92\,\alpha_{\rm EM} / (2\pi f)$ to convert between $g_{\phi\gamma\gamma}$ and $1/f$.}. Interestingly, GW experiments are most sensitive for large values of the decay constant $f$ corresponding to very weakly coupled axions. These probes are therefore highly complementary to other existing limits (orange shaded) or planned searches (orange lines), which are typically more sensitive for larger couplings. 
An exception is the constraint coming from black hole superradiance (gray shaded), which is also most reliable for large decay constant $f$ and also indirectly relies on GW observations~\cite{Cardoso:2018tly,Arvanitaki:2014wva}. It should also be emphasized that the GW signal regions do not depend on the axion relic abundance today, and therefore do not require the axion to account for all of DM. 
The non-decoupling behavior of the GW signal is due to the fact that larger $f$ corresponds to more energy in the axion field $\Omega_{\phi}^{\rm osc} \propto m^{2} \theta^{2}f^{2}$ which is available to be converted into gravitational radiation. This holds as long as the initial misalignment angle $\theta$ takes on natural values of $\mathcal{O}(1)$~\footnote{Additionally, we are always assuming $m_{X} \lesssim m/2$ and  $\alpha \sim 10-100$ such that the particle production process is efficient, see e.g. Refs.~\cite{Agrawal:2017eqm,Machado:2018nqk}.}. Indeed, for our numerical results here, $\theta=1$ is chosen.

\begin{figure}[t!]
\centering 
\includegraphics[width=0.95\columnwidth]{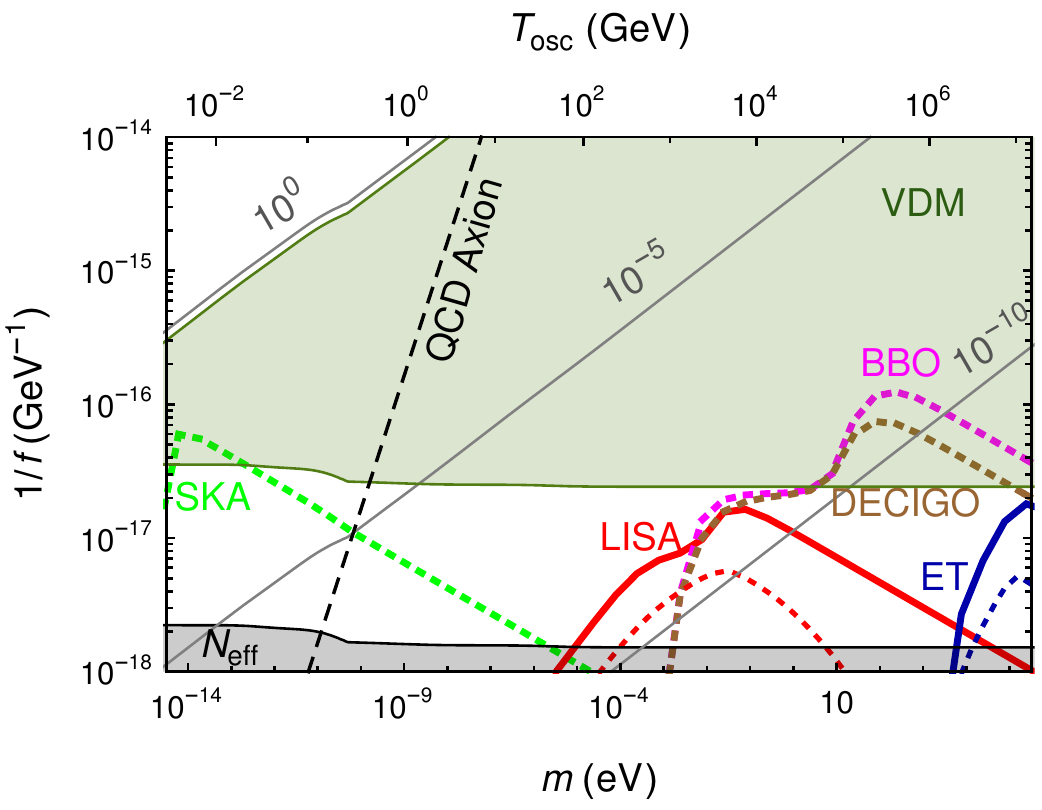}
\caption{Cosmological constraints on the model. The green shaded region indicates where dark photons could be vector dark matter (VDM), while the region labeled $N_{\rm eff}$ is excluded by constraints on the number of relativistic degrees of freedom. Furthermore, the required suppression of the axion abundance is indicated by the diagonal gray lines, in order not to overproduce axion DM. As before, the colored curves show the regions accessible to future GW experiments. In addition, we also show the region where LISA (dashed red) or ET (dashed blue) can detect the chirality of the GW signal. }
\label{fig:bounds}
\end{figure}

In \cref{fig:bounds}, we show a close up of the parameter space that leads to detectable signals, as well as bounds arising from cosmology. If the dark photon stays relativistic until recombination, the number of effective relativistic degrees of freedom $N_{\rm eff}$ sets an upper bound on the decay constant $f$.  A simple estimate can be done assuming that all the energy in the axion field is converted into dark gauge bosons. This leads to a bound of $f \lesssim (5-7)\times 10^{17}$ GeV shown in \cref{fig:bounds}, depending on whether the axion starts oscillating before or after the QCD phase transition. 
The dark photon might also become non-relativistic before recombination and therefore contribute to DM, as in Refs.~\cite{Dror:2018pdh,Co:2018lka,Bastero-Gil:2018uel,Agrawal:2018vin}.  The shaded green region of \cref{fig:bounds} shows the potential parameter space for vector DM (VDM), which is cut off at the lower bound of $f\approx 3\times 10^{16}$ GeV where the dark photons are too hot to be compatible with structure formation. Additional information on the derivation of these bounds can be found in~\cref{app:relics}. The diagonal gray lines indicate how much the axion abundance must be suppressed compared to the ordinary misalignment case to avoid over-production~\footnote{While the exponential production of dark photons can lead to a very strong suppression~\cite{Agrawal:2017eqm,Machado:2018nqk}, it is currently unclear whether a suppression by more than three orders of magnitude persists if backreaction effects are taken into account~\cite{Kitajima:2017peg}. Otherwise, other mechanisms for depleting the axion abundance might be needed in the bottom right region of parameter space.\vspace{0.75mm}}.

A smoking gun for Audible Axion models is the completely chiral nature of the peak of the GW spectrum, inherited from the parity violation in the dark photon population~\cite{Machado:2018nqk}. This can provide powerful background rejection, since SGWBs from astrophysical sources are not expected to carry a net polarization. 
It has been pointed out that the dipolar anisotropy induced by the Doppler shift due to the relative motion of our solar system with respect to the cosmic reference frame can be exploited to allow planar detectors to detect net circular polarization~\cite{Seto:2007tn,Seto:2008sr,Smith:2016jqs,Seto:2006hf,Seto:2006dz}.  In particular, LISA and ET would be able to detect net circular polarization with an SNR of $\mathcal{O}(1)$ for a SGWB with amplitude $h^{2} \Omega_{\rm GW} \sim 10^{-11}$~\cite{Domcke:2019zls}. In \cref{fig:bounds}, we indicate using dashed lines the region in parameter space where the signal is strong enough such that LISA and ET can pick up on the polarization following the analysis of Ref.~\cite{Domcke:2019zls}. Of course, if a network of non-coplanar detectors is available in a particular frequency range, GW polarization can be detected without paying the $\mathcal{O}(10^{-3})$ suppression factor due to our peculiar velocity~\cite{Seto:2007tn,Crowder:2012ik}.

\section{Discussion and Conclusions}

In~\cite{Machado:2018nqk} we showed that a SGWB can be produced by an axion coupled to a dark photon, specifically by the tachyonic instability induced in the dark photon by the axion dynamics. This instability leads to exponential growth of dark photon vacuum fluctuations which act as the GW source. Here, we have shown that this GW signal is also produced for a broader class of models that allow for a massive dark photon and/or kinetic mixing with the SM photon. Furthermore, we argue that couplings of the axion to gluons and photons do not affect the success of the mechanism, which illustrates the viability of the QCD axion case.

The central results of our paper are \cref{fig:wolframs_money} and \cref{fig:bounds}, where we show the regions of axion parameter space that may be probed by future GW experiments. Since the GW signal is strongest for large decay constants, GWs probe complementary regions of parameter space to most other experiments relying on couplings of the axion to the visible sector (proportional to the inverse of the decay constant) that are sizeable. In \cref{fig:bounds}, we zoom in on the GW signal region and show cosmological constraints as well as the region where the dark photon itself could be DM. For most of the parameter space relevant for GW detectors, the axion relic abundance needs to be strongly suppressed, which might require an extension of the model. 

Since the shape of the GW signal is universal for dark photon masses less than roughly the axion mass, we provide a fit template function which parameterizes the dependence of the GW amplitude and peak frequency on the axion mass $m$ and decay constant $f$. This fit, which is extracted from our new simulation with an order of magnitude higher mode density, provides a quick translation from the underlying model parameters to the detectability of the GW signal for all experiments and parameter points, providing a tool which can be directly used by experimental collaborations to probe our and similar models. 

\subsection*{Acknowledgments}
We would like to thank Masha Baryakhtar for illuminating discussions regarding the black hole superradiance constraints. 
The work of CSM was supported by the Alexander von Humboldt Foundation, in the framework of the Sofja Kovalevskaja Award 2016, endowed by the German Federal Ministry of Education and Research. Work of WR was partially funded by the Deutsche Forschungsgemeinschaft (DFG), Project ID 438947057. We also acknowledge support by  the  Cluster  of  Excellence  ``Precision  Physics,  Fundamental
Interactions, and Structure of Matter" (PRISMA$^+$ EXC 2118/1) funded by the German Research Foundation (DFG) within the German Excellence Strategy (Project ID 39083149).

\bibliographystyle{JHEP}
\bibliography{draft.bib}

\providecommand{\href}[2]{#2}\begingroup\raggedright\begin{thebibliography}{10}

\bibitem{PhysRevLett.38.1440}
R.~D. Peccei and H.~R. Quinn, {\it $\mathrm{CP}$ conservation in the presence
  of pseudoparticles},  {\em Phys. Rev. Lett.} {\bf 38} (Jun, 1977) 1440--1443.

\bibitem{Graham:2015cka}
P.~W. Graham, D.~E. Kaplan, and S.~Rajendran, {\it {Cosmological Relaxation of
  the Electroweak Scale}},  {\em Phys. Rev. Lett.} {\bf 115} (2015), no.~22
  221801, [\href{http://arxiv.org/abs/1504.07551}{{\tt arXiv:1504.07551}}].

\bibitem{PhysRevLett.65.3233}
K.~Freese, J.~A. Frieman, and A.~V. Olinto, {\it Natural inflation with pseudo
  nambu-goldstone bosons},  {\em Phys. Rev. Lett.} {\bf 65} (Dec, 1990)
  3233--3236.

\bibitem{Preskill:1982cy}
J.~Preskill, M.~B. Wise, and F.~Wilczek, {\it {Cosmology of the Invisible
  Axion}},  {\em Phys. Lett.} {\bf 120B} (1983) 127--132.

\bibitem{Abbott:1982af}
L.~F. Abbott and P.~Sikivie, {\it {A Cosmological Bound on the Invisible
  Axion}},  {\em Phys. Lett.} {\bf 120B} (1983) 133--136.

\bibitem{Dine:1982ah}
M.~Dine and W.~Fischler, {\it {The Not So Harmless Axion}},  {\em Phys. Lett.}
  {\bf 120B} (1983) 137--141.

\bibitem{Arvanitaki:2009fg}
A.~Arvanitaki, S.~Dimopoulos, S.~Dubovsky, N.~Kaloper, and J.~March-Russell,
  {\it {String Axiverse}},  {\em Phys. Rev.} {\bf D81} (2010) 123530,
  [\href{http://arxiv.org/abs/0905.4720}{{\tt arXiv:0905.4720}}].

\bibitem{Machado:2018nqk}
C.~S. Machado, W.~Ratzinger, P.~Schwaller, and B.~A. Stefanek, {\it {Audible
  Axions}},  {\em JHEP} {\bf 01} (2019) 053,
  [\href{http://arxiv.org/abs/1811.01950}{{\tt arXiv:1811.01950}}].

\bibitem{Domcke:2019zls}
V.~Domcke, J.~Garcia-Bellido, M.~Peloso, M.~Pieroni, A.~Ricciardone, L.~Sorbo,
  and G.~Tasinato, {\it {Measuring the net circular polarization of the
  stochastic gravitational wave background with interferometers}},
  \href{http://arxiv.org/abs/1910.08052}{{\tt arXiv:1910.08052}}.

\bibitem{Agrawal:2017cmd}
P.~Agrawal, J.~Fan, M.~Reece, and L.-T. Wang, {\it {Experimental Targets for
  Photon Couplings of the QCD Axion}},  {\em JHEP} {\bf 02} (2018) 006,
  [\href{http://arxiv.org/abs/1709.06085}{{\tt arXiv:1709.06085}}].

\bibitem{Ratra:1991bn}
B.~Ratra, {\it {Cosmological 'seed' magnetic field from inflation}},  {\em
  Astrophys. J.} {\bf 391} (1992) L1--L4.

\bibitem{Garretson:1992vt}
W.~D. Garretson, G.~B. Field, and S.~M. Carroll, {\it {Primordial magnetic
  fields from pseudoGoldstone bosons}},  {\em Phys. Rev.} {\bf D46} (1992)
  5346--5351, [\href{http://arxiv.org/abs/hep-ph/9209238}{{\tt
  hep-ph/9209238}}].

\bibitem{Field:1998hi}
G.~B. Field and S.~M. Carroll, {\it {Cosmological magnetic fields from
  primordial helicity}},  {\em Phys. Rev.} {\bf D62} (2000) 103008,
  [\href{http://arxiv.org/abs/astro-ph/9811206}{{\tt astro-ph/9811206}}].

\bibitem{Lee:2001hj}
D.-S. Lee, W.-l. Lee, and K.-W. Ng, {\it {Primordial magnetic fields from dark
  energy}},  {\em Phys. Lett.} {\bf B542} (2002) 1--7,
  [\href{http://arxiv.org/abs/astro-ph/0109184}{{\tt astro-ph/0109184}}].

\bibitem{Campanelli:2005ye}
L.~Campanelli and M.~Giannotti, {\it {Magnetic helicity generation from the
  cosmic axion field}},  {\em Phys. Rev.} {\bf D72} (2005) 123001,
  [\href{http://arxiv.org/abs/astro-ph/0508653}{{\tt astro-ph/0508653}}].

\bibitem{Anber:2006xt}
M.~M. Anber and L.~Sorbo, {\it {N-flationary magnetic fields}},  {\em JCAP}
  {\bf 0610} (2006) 018, [\href{http://arxiv.org/abs/astro-ph/0606534}{{\tt
  astro-ph/0606534}}].

\bibitem{Anber:2009ua}
M.~M. Anber and L.~Sorbo, {\it {Naturally inflating on steep potentials through
  electromagnetic dissipation}},  {\em Phys. Rev.} {\bf D81} (2010) 043534,
  [\href{http://arxiv.org/abs/0908.4089}{{\tt arXiv:0908.4089}}].

\bibitem{Barnaby:2010vf}
N.~Barnaby and M.~Peloso, {\it {Large Nongaussianity in Axion Inflation}},
  {\em Phys. Rev. Lett.} {\bf 106} (2011) 181301,
  [\href{http://arxiv.org/abs/1011.1500}{{\tt arXiv:1011.1500}}].

\bibitem{Barnaby:2011vw}
N.~Barnaby, R.~Namba, and M.~Peloso, {\it {Phenomenology of a Pseudo-Scalar
  Inflaton: Naturally Large Nongaussianity}},  {\em JCAP} {\bf 1104} (2011)
  009, [\href{http://arxiv.org/abs/1102.4333}{{\tt arXiv:1102.4333}}].

\bibitem{Barnaby:2011qe}
N.~Barnaby, E.~Pajer, and M.~Peloso, {\it {Gauge Field Production in Axion
  Inflation: Consequences for Monodromy, non-Gaussianity in the CMB, and
  Gravitational Waves at Interferometers}},  {\em Phys. Rev.} {\bf D85} (2012)
  023525, [\href{http://arxiv.org/abs/1110.3327}{{\tt arXiv:1110.3327}}].

\bibitem{Barnaby:2012tk}
N.~Barnaby, R.~Namba, and M.~Peloso, {\it {Observable non-gaussianity from
  gauge field production in slow roll inflation, and a challenging connection
  with magnetogenesis}},  {\em Phys. Rev.} {\bf D85} (2012) 123523,
  [\href{http://arxiv.org/abs/1202.1469}{{\tt arXiv:1202.1469}}].

\bibitem{Adshead:2013qp}
P.~Adshead, E.~Martinec, and M.~Wyman, {\it {Gauge fields and inflation: Chiral
  gravitational waves, fluctuations, and the Lyth bound}},  {\em Phys. Rev.}
  {\bf D88} (2013), no.~2 021302, [\href{http://arxiv.org/abs/1301.2598}{{\tt
  arXiv:1301.2598}}].

\bibitem{Adshead:2015pva}
P.~Adshead, J.~T. Giblin, T.~R. Scully, and E.~I. Sfakianakis, {\it
  {Gauge-preheating and the end of axion inflation}},  {\em JCAP} {\bf 1512}
  (2015), no.~12 034, [\href{http://arxiv.org/abs/1502.06506}{{\tt
  arXiv:1502.06506}}].

\bibitem{Giblin:2017wlo}
J.~T. Giblin, G.~Kane, E.~Nesbit, S.~Watson, and Y.~Zhao, {\it {Was the
  Universe Actually Radiation Dominated Prior to Nucleosynthesis?}},  {\em
  Phys. Rev.} {\bf D96} (2017), no.~4 043525,
  [\href{http://arxiv.org/abs/1706.08536}{{\tt arXiv:1706.08536}}].

\bibitem{Hook:2016mqo}
A.~Hook and G.~Marques-Tavares, {\it {Relaxation from particle production}},
  {\em JHEP} {\bf 12} (2016) 101, [\href{http://arxiv.org/abs/1607.01786}{{\tt
  arXiv:1607.01786}}].

\bibitem{Domcke:2016bkh}
V.~Domcke, M.~Pieroni, and P.~Bin{\'e}truy, {\it {Primordial gravitational
  waves for universality classes of pseudoscalar inflation}},  {\em JCAP} {\bf
  1606} (2016) 031, [\href{http://arxiv.org/abs/1603.01287}{{\tt
  arXiv:1603.01287}}].

\bibitem{Kitajima:2017peg}
N.~Kitajima, T.~Sekiguchi, and F.~Takahashi, {\it {Cosmological abundance of
  the QCD axion coupled to hidden photons}},  {\em Phys. Lett.} {\bf B781}
  (2018) 684--687, [\href{http://arxiv.org/abs/1711.06590}{{\tt
  arXiv:1711.06590}}].

\bibitem{Agrawal:2017eqm}
P.~Agrawal, G.~Marques-Tavares, and W.~Xue, {\it {Opening up the QCD axion
  window}},  {\em JHEP} {\bf 03} (2018) 049,
  [\href{http://arxiv.org/abs/1708.05008}{{\tt arXiv:1708.05008}}].

\bibitem{Fonseca:2018xzp}
N.~Fonseca, E.~Morgante, and G.~Servant, {\it {Higgs relaxation after
  inflation}},  {\em JHEP} {\bf 10} (2018) 020,
  [\href{http://arxiv.org/abs/1805.04543}{{\tt arXiv:1805.04543}}].

\bibitem{Dror:2018pdh}
J.~A. Dror, K.~Harigaya, and V.~Narayan, {\it {Parametric Resonance Production
  of Ultralight Vector Dark Matter}},
  \href{http://arxiv.org/abs/1810.07195}{{\tt arXiv:1810.07195}}.

\bibitem{Co:2018lka}
R.~T. Co, A.~Pierce, Z.~Zhang, and Y.~Zhao, {\it {Dark Photon Dark Matter
  Produced by Axion Oscillations}},
  \href{http://arxiv.org/abs/1810.07196}{{\tt arXiv:1810.07196}}.

\bibitem{Bastero-Gil:2018uel}
M.~Bastero-Gil, J.~Santiago, L.~Ubaldi, and R.~Vega-Morales, {\it {Vector dark
  matter production at the end of inflation}},
  \href{http://arxiv.org/abs/1810.07208}{{\tt arXiv:1810.07208}}.

\bibitem{Agrawal:2018vin}
P.~Agrawal, N.~Kitajima, M.~Reece, T.~Sekiguchi, and F.~Takahashi, {\it {Relic
  Abundance of Dark Photon Dark Matter}},
  \href{http://arxiv.org/abs/1810.07188}{{\tt arXiv:1810.07188}}.

\bibitem{Soda:2017dsu}
J.~Soda and Y.~Urakawa, {\it {Cosmological imprints of string axions in
  plateau}},  {\em Eur. Phys. J.} {\bf C78} (2018), no.~9 779,
  [\href{http://arxiv.org/abs/1710.00305}{{\tt arXiv:1710.00305}}].

\bibitem{Kitajima:2018zco}
N.~Kitajima, J.~Soda, and Y.~Urakawa, {\it {Gravitational wave forest from
  string axiverse}},  {\em JCAP} {\bf 1810} (2018), no.~10 008,
  [\href{http://arxiv.org/abs/1807.07037}{{\tt arXiv:1807.07037}}].

\bibitem{Carenza:2019vzg}
P.~Carenza, A.~Mirizzi, and G.~Sigl, {\it {Dynamical evolution of axion
  condensates under stimulated decays into photons}},
  \href{http://arxiv.org/abs/1911.07838}{{\tt arXiv:1911.07838}}.

\bibitem{Alonso-Alvarez:2019ssa}
G.~Alonso-{\'A}lvarez, R.~S. Gupta, J.~Jaeckel, and M.~Spannowsky, {\it {On the
  Wondrous Stability of ALP Dark Matter}},
  \href{http://arxiv.org/abs/1911.07885}{{\tt arXiv:1911.07885}}.

\bibitem{Adshead:2018doq}
P.~Adshead, J.~T. Giblin, and Z.~J. Weiner, {\it {Gravitational waves from
  gauge preheating}},  {\em Phys. Rev.} {\bf D98} (2018), no.~4 043525,
  [\href{http://arxiv.org/abs/1805.04550}{{\tt arXiv:1805.04550}}].

\bibitem{Adshead:2019igv}
P.~Adshead, J.~T. Giblin, M.~Pieroni, and Z.~J. Weiner, {\it {Constraining
  axion inflation with gravitational waves across 29 decades in frequency}},
  \href{http://arxiv.org/abs/1909.12843}{{\tt arXiv:1909.12843}}.

\bibitem{Adshead:2019lbr}
P.~Adshead, J.~T. Giblin, M.~Pieroni, and Z.~J. Weiner, {\it {Constraining
  axion inflation with gravitational waves from preheating}},
  \href{http://arxiv.org/abs/1909.12842}{{\tt arXiv:1909.12842}}.

\bibitem{Holdom:1985ag}
B.~Holdom, {\it {Two U(1)'s and Epsilon Charge Shifts}},  {\em Phys. Lett.}
  {\bf 166B} (1986) 196--198.

\bibitem{Dubovsky:2015cca}
S.~Dubovsky and G.~Hern{\'a}ndez-Chifflet, {\it {Heating up the Galaxy with
  Hidden Photons}},  {\em JCAP} {\bf 1512} (2015), no.~12 054,
  [\href{http://arxiv.org/abs/1509.00039}{{\tt arXiv:1509.00039}}].

\bibitem{Jaeckel:2010ni}
J.~Jaeckel and A.~Ringwald, {\it {The Low-Energy Frontier of Particle
  Physics}},  {\em Ann. Rev. Nucl. Part. Sci.} {\bf 60} (2010) 405--437,
  [\href{http://arxiv.org/abs/1002.0329}{{\tt arXiv:1002.0329}}].

\bibitem{Jaeckel:2007ch}
J.~Jaeckel and A.~Ringwald, {\it {A Cavity Experiment to Search for Hidden
  Sector Photons}},  {\em Phys. Lett.} {\bf B659} (2008) 509--514,
  [\href{http://arxiv.org/abs/0707.2063}{{\tt arXiv:0707.2063}}].

\bibitem{Abel:2006qt}
S.~A. Abel, J.~Jaeckel, V.~V. Khoze, and A.~Ringwald, {\it {Illuminating the
  Hidden Sector of String Theory by Shining Light through a Magnetic Field}},
  {\em Phys. Lett.} {\bf B666} (2008) 66--70,
  [\href{http://arxiv.org/abs/hep-ph/0608248}{{\tt hep-ph/0608248}}].

\bibitem{Abel:2008ai}
S.~A. Abel, M.~D. Goodsell, J.~Jaeckel, V.~V. Khoze, and A.~Ringwald, {\it
  {Kinetic Mixing of the Photon with Hidden U(1)s in String Phenomenology}},
  {\em JHEP} {\bf 07} (2008) 124, [\href{http://arxiv.org/abs/0803.1449}{{\tt
  arXiv:0803.1449}}].

\bibitem{Ahlers:2008qc}
M.~Ahlers, J.~Jaeckel, J.~Redondo, and A.~Ringwald, {\it {Probing Hidden Sector
  Photons through the Higgs Window}},  {\em Phys. Rev.} {\bf D78} (2008)
  075005, [\href{http://arxiv.org/abs/0807.4143}{{\tt arXiv:0807.4143}}].

\bibitem{Redondo:2008en}
J.~Redondo, {\it {Bounds on Very Weakly Interacting Sub-eV Particles (WISPs)
  from Cosmology and Astrophysics}},  in {\em {Proceedings, 4th Patras Workshop
  on Axions, WIMPs and WISPs (AXION-WIMP 2008): Hamburg, Germany, June 18-21,
  2008}}, pp.~23--26, 2008.
\newblock \href{http://arxiv.org/abs/0810.3200}{{\tt arXiv:0810.3200}}.

\bibitem{Redondo:2008aa}
J.~Redondo, {\it {Helioscope Bounds on Hidden Sector Photons}},  {\em JCAP}
  {\bf 0807} (2008) 008, [\href{http://arxiv.org/abs/0801.1527}{{\tt
  arXiv:0801.1527}}].

\bibitem{Mirizzi:2009iz}
A.~Mirizzi, J.~Redondo, and G.~Sigl, {\it {Microwave Background Constraints on
  Mixing of Photons with Hidden Photons}},  {\em JCAP} {\bf 0903} (2009) 026,
  [\href{http://arxiv.org/abs/0901.0014}{{\tt arXiv:0901.0014}}].

\bibitem{Essig:2009nc}
R.~Essig, P.~Schuster, and N.~Toro, {\it {Probing Dark Forces and Light Hidden
  Sectors at Low-Energy e+e- Colliders}},  {\em Phys. Rev.} {\bf D80} (2009)
  015003, [\href{http://arxiv.org/abs/0903.3941}{{\tt arXiv:0903.3941}}].

\bibitem{Batell:2009di}
B.~Batell, M.~Pospelov, and A.~Ritz, {\it {Exploring Portals to a Hidden Sector
  Through Fixed Targets}},  {\em Phys. Rev.} {\bf D80} (2009) 095024,
  [\href{http://arxiv.org/abs/0906.5614}{{\tt arXiv:0906.5614}}].

\bibitem{Afanasev:2008fv}
A.~Afanasev, O.~K. Baker, K.~B. Beard, G.~Biallas, J.~Boyce, M.~Minarni,
  R.~Ramdon, M.~Shinn, and P.~Slocum, {\it {New Experimental Limit on Photon
  Hidden-Sector Paraphoton Mixing}},  {\em Phys. Lett.} {\bf B679} (2009)
  317--320, [\href{http://arxiv.org/abs/0810.4189}{{\tt arXiv:0810.4189}}].

\bibitem{Jaeckel:2008sz}
J.~Jaeckel and J.~Redondo, {\it {Searching Hidden-sector Photons inside a
  Superconducting Box}},  {\em EPL} {\bf 84} (2008), no.~3 31002,
  [\href{http://arxiv.org/abs/0806.1115}{{\tt arXiv:0806.1115}}].

\bibitem{Gninenko:2008pz}
S.~N. Gninenko and J.~Redondo, {\it {On search for eV hidden sector photons in
  Super-Kamiokande and CAST experiments}},  {\em Phys. Lett.} {\bf B664} (2008)
  180--184, [\href{http://arxiv.org/abs/0804.3736}{{\tt arXiv:0804.3736}}].

\bibitem{Baryakhtar:2018doz}
M.~Baryakhtar, J.~Huang, and R.~Lasenby, {\it {Axion and hidden photon dark
  matter detection with multilayer optical haloscopes}},  {\em Phys. Rev.} {\bf
  D98} (2018), no.~3 035006, [\href{http://arxiv.org/abs/1803.11455}{{\tt
  arXiv:1803.11455}}].

\bibitem{PhysRevLett.43.103}
J.~E. Kim, {\it Weak-interaction singlet and strong $\mathrm{CP}$ invariance},
  {\em Phys. Rev. Lett.} {\bf 43} (Jul, 1979) 103--107.

\bibitem{SHIFMAN1980493}
M.~Shifman, A.~Vainshtein, and V.~Zakharov, {\it Can confinement ensure natural
  cp invariance of strong interactions?},  {\em Nuclear Physics B} {\bf 166}
  (1980), no.~3 493 -- 506.

\bibitem{Kapusta:2006pm}
J.~I. Kapusta and C.~Gale, {\em {Finite-temperature field theory: Principles
  and applications}}.
\newblock Cambridge Monographs on Mathematical Physics. Cambridge University
  Press, 2011.

\bibitem{Kraemmer:1995qe}
U.~Kraemmer, A.~K. Rebhan, and H.~Schulz, {\it {Hot scalar electrodynamics as a
  toy model for hot QCD}},  in {\em {From thermal field theory to neural
  networks: A day to remember Tanguy Altherr. Proceedings, Meeting, Geneva,
  Switzerland, November 4, 1994}}, pp.~13--23, 1995.
\newblock \href{http://arxiv.org/abs/hep-ph/9505307}{{\tt hep-ph/9505307}}.

\bibitem{LINDE1980289}
A.~Linde, {\it Infrared problem in the thermodynamics of the yang-mills gas},
  {\em Physics Letters B} {\bf 96} (1980), no.~3 289 -- 292.

\bibitem{RevModPhys.53.43}
D.~J. Gross, R.~D. Pisarski, and L.~G. Yaffe, {\it Qcd and instantons at finite
  temperature},  {\em Rev. Mod. Phys.} {\bf 53} (Jan, 1981) 43--80.

\bibitem{Espinosa:1992kf}
J.~R. Espinosa, M.~Quiros, and F.~Zwirner, {\it {On the nature of the
  electroweak phase transition}},  {\em Phys. Lett.} {\bf B314} (1993)
  206--216, [\href{http://arxiv.org/abs/hep-ph/9212248}{{\tt hep-ph/9212248}}].

\bibitem{Breitbach:2018ddu}
M.~Breitbach, J.~Kopp, E.~Madge, T.~Opferkuch, and P.~Schwaller, {\it {Dark,
  Cold, and Noisy: Constraining Secluded Hidden Sectors with Gravitational
  Waves}},  {\em JCAP} {\bf 1907} (2019), no.~07 007,
  [\href{http://arxiv.org/abs/1811.11175}{{\tt arXiv:1811.11175}}].

\bibitem{Cardoso:2018tly}
V.~Cardoso, O.~J.~C. Dias, G.~S. Hartnett, M.~Middleton, P.~Pani, and J.~E.
  Santos, {\it {Constraining the mass of dark photons and axion-like particles
  through black-hole superradiance}},  {\em JCAP} {\bf 1803} (2018), no.~03
  043, [\href{http://arxiv.org/abs/1801.01420}{{\tt arXiv:1801.01420}}].

\bibitem{Arvanitaki:2014wva}
A.~Arvanitaki, M.~Baryakhtar, and X.~Huang, {\it {Discovering the QCD Axion
  with Black Holes and Gravitational Waves}},  {\em Phys. Rev.} {\bf D91}
  (2015), no.~8 084011, [\href{http://arxiv.org/abs/1411.2263}{{\tt
  arXiv:1411.2263}}].

\bibitem{Seto:2007tn}
N.~Seto and A.~Taruya, {\it {Measuring a Parity Violation Signature in the
  Early Universe via Ground-based Laser Interferometers}},  {\em Phys. Rev.
  Lett.} {\bf 99} (2007) 121101, [\href{http://arxiv.org/abs/0707.0535}{{\tt
  arXiv:0707.0535}}].

\bibitem{Seto:2008sr}
N.~Seto and A.~Taruya, {\it {Polarization analysis of gravitational-wave
  backgrounds from the correlation signals of ground-based interferometers:
  Measuring a circular-polarization mode}},  {\em Phys. Rev.} {\bf D77} (2008)
  103001, [\href{http://arxiv.org/abs/0801.4185}{{\tt arXiv:0801.4185}}].

\bibitem{Smith:2016jqs}
T.~L. Smith and R.~Caldwell, {\it {Sensitivity to a Frequency-Dependent
  Circular Polarization in an Isotropic Stochastic Gravitational Wave
  Background}},  {\em Phys. Rev.} {\bf D95} (2017), no.~4 044036,
  [\href{http://arxiv.org/abs/1609.05901}{{\tt arXiv:1609.05901}}].

\bibitem{Seto:2006hf}
N.~Seto, {\it {Prospects for direct detection of circular polarization of
  gravitational-wave background}},  {\em Phys. Rev. Lett.} {\bf 97} (2006)
  151101, [\href{http://arxiv.org/abs/astro-ph/0609504}{{\tt
  astro-ph/0609504}}].

\bibitem{Seto:2006dz}
N.~Seto, {\it {Quest for circular polarization of gravitational wave background
  and orbits of laser interferometers in space}},  {\em Phys. Rev.} {\bf D75}
  (2007) 061302, [\href{http://arxiv.org/abs/astro-ph/0609633}{{\tt
  astro-ph/0609633}}].

\bibitem{Crowder:2012ik}
S.~Crowder, R.~Namba, V.~Mandic, S.~Mukohyama, and M.~Peloso, {\it {Measurement
  of Parity Violation in the Early Universe using Gravitational-wave
  Detectors}},  {\em Phys. Lett. B} {\bf 726} (2013) 66--71,
  [\href{http://arxiv.org/abs/1212.4165}{{\tt arXiv:1212.4165}}].

\bibitem{Aghanim:2018eyx}
{\bf Planck} Collaboration, N.~Aghanim et~al., {\it {Planck 2018 results. VI.
  Cosmological parameters}},  \href{http://arxiv.org/abs/1807.06209}{{\tt
  arXiv:1807.06209}}.

\bibitem{Narayanan:2000tp}
V.~K. Narayanan, D.~N. Spergel, R.~Dave, and C.-P. Ma, {\it {Constraints on the
  mass of warm dark matter particles and the shape of the linear power spectrum
  from the Ly$\alpha$ forest}},  {\em Astrophys. J.} {\bf 543} (2000)
  L103--L106, [\href{http://arxiv.org/abs/astro-ph/0005095}{{\tt
  astro-ph/0005095}}].

\bibitem{Hansen:2001zv}
S.~H. Hansen, J.~Lesgourgues, S.~Pastor, and J.~Silk, {\it {Constraining the
  window on sterile neutrinos as warm dark matter}},  {\em Mon. Not. Roy.
  Astron. Soc.} {\bf 333} (2002) 544--546,
  [\href{http://arxiv.org/abs/astro-ph/0106108}{{\tt astro-ph/0106108}}].

\end{thebibliography}\endgroup
	
\appendix

\section{Effects of Finite Dark Photon Mass}
\label{app:finiteMass}
Here, we discuss the impact of non-zero mass for the dark photon. The dark gauge boson dispersion relation for finite mass is
\begin{equation}
 \omega^{2}_{\pm} = k^{2} +a^{2}m_{X}^{2} \mp k \frac{\alpha}{f}\phi' \,.
 \label{eq:ppb}
\end{equation}
The mode $\tilde{k}$ that minimizes the frequency is $\tilde{k} = \alpha\phi' /(2f)$, so no particle production is allowed unless
\begin{equation}
 |\phi'| > \frac{2 f a m_{X}}{\alpha} \, .
 \label{eq:ppb}
\end{equation}
We call the minimum speed at which particle production becomes kinematically allowed $|\phi'_{\rm min}| \equiv 2 f a m_{X}/\alpha$. Tachyonic particle production will end when $\phi'$ drops below this value, at which point the most tachyonic scale is
\begin{equation}
\tilde{k}_{\rm min} =\frac{|\phi'_{\rm min}|}{2f} = a m_{X} \, .
\label{eq:firstK}
\end{equation}
\begin{figure}[h]
\centering
\includegraphics[width=0.95\columnwidth]{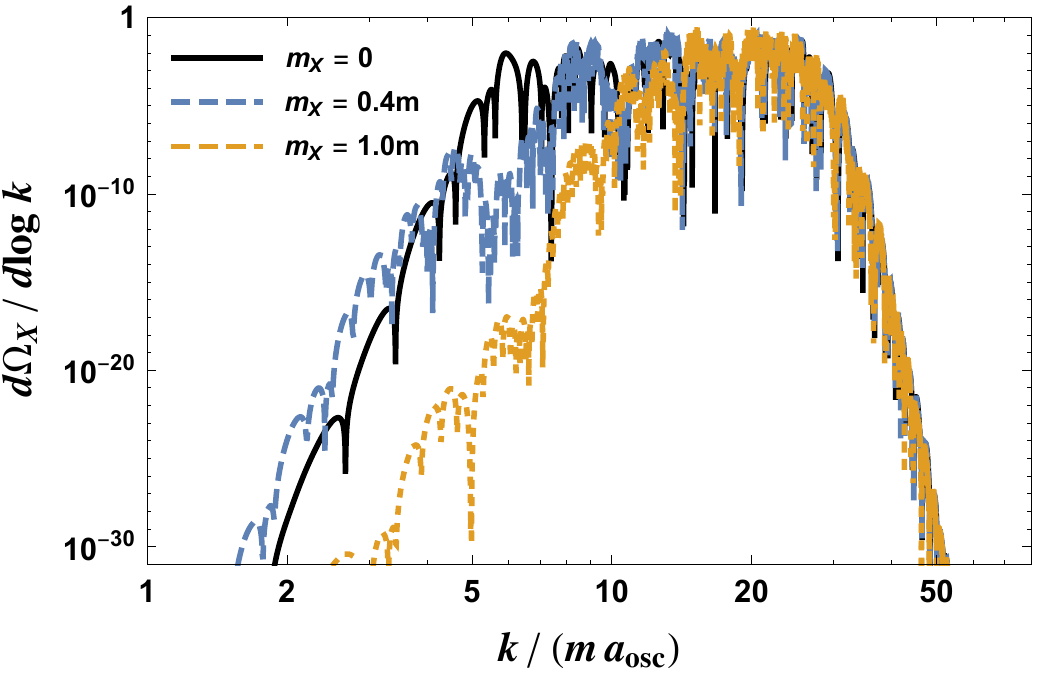}
\caption{Dark photon spectral energy density for different values of the dark photon mass $m_{X}$ using the improved numerical simulation for the ALP2 benchmark point of Ref.~\cite{Machado:2018nqk}.}
\label{fig:GPspecMx}
\end{figure}
Before the backreaction from particle production becomes strong, $\phi'$ scales roughly as $\phi' \approx \theta f (a_{\rm osc} / a)^{3/2} a m$, so we can estimate the value of the scale factor at the time when kinematics shuts off the particle production process as
\begin{equation}
 \frac{a}{a_{\rm osc}} = \left( \frac{\alpha\theta}{2} \frac{m}{m_{X}} \right)^{2/3} \,,
\end{equation}
at which time the most tachyonic scale is 
\begin{equation}
\frac{\tilde{k}_{\rm min}}{m a_{\rm osc}} = \frac{a}{a_{\rm osc}} \frac{m_{X}}{m} =  \left( \frac{\alpha\theta}{2}\right)^{2/3} \left(\frac{m_{X}}{m}\right)^{1/3} \, .
\label{eq:GPcutoff}
\end{equation}
We expect this scale to provide a hard low-$k$ cutoff in the gauge power spectrum. Indeed, we show in \cref{fig:GPspecMx} the dark photon power spectrum for different values $m_{X}$, where \cref{eq:GPcutoff} provides a good description of the low-$k$ cutoff.
\subsection{Tachyonic Band}
Following the analysis of Ref.~\cite{Machado:2018nqk}, the tachyonic band $k_{-} < k < k_{+}$ is found by solving for the tachyonic modes with growth timescales less than the conformal oscillation time. Specifically, it is given by solving $\omega^{2} = -(am)^{2}$, with the result
\begin{equation}
k_{\pm} = \tilde{k} \left[ 1\pm \sqrt{1-\left( \frac{2}{\alpha\theta}\right)^{2} \left( \frac{a}{a_{\rm osc}}\right)^{3}\left( 1+ \frac{m_{X}^{2}}{m^{2}}\right)}\, \right] \,,
\end{equation}
which reproduces the massless result when $m_{X} \rightarrow 0$. The band closes once the scale factor has increased by an amount
\begin{equation}
 \frac{a}{a_{\rm osc}} = \left( \frac{\alpha\theta}{2} \right)^{2/3}\left( 1+ \frac{m_{X}^{2}}{m^{2}}\right)^{-1/3} \,.
\end{equation}
Because the scale factor for tachyonic band closure approaches the kinematic closure value for $m_{X} / m \gg 1$, one can see that the tachyonic band always closes before kinematics shut off the tachyonic production. The most tachyonic scale at the time of tachyonic band closure is 
\begin{equation}
\frac{\tilde{k}}{m a_{\rm osc}} =  \left( \frac{\alpha\theta}{2}\right)^{2/3} \left( 1+ \frac{m_{X}^{2}}{m^{2}}\right)^{1/6} \, ,
\label{eq:GPeffCO}
\end{equation}
which explains why the spectrum starts to fall off before the hard kinematic cutoff given by \cref{eq:GPcutoff}, since modes below this scale are not efficiently produced.
\subsection{Gravitational Wave Spectrum}
In the limit $\alpha\theta \gtrsim 10$, requiring $a/a_{\rm osc} > 1$ in both \cref{eq:GPcutoff,eq:GPeffCO} gives the condition
\begin{equation}
m_{X} \lesssim \frac{\alpha \theta}{2} m \,,
\end{equation}
that is required in order to have any tachyonic production. However, to keep the tachyonic band open until the scale factor has grown by an order of magnitude (as is typically required to produce an observable GW signal), we require
\begin{equation}
m_{X} \lesssim \frac{\alpha \theta}{100} m \,,
\end{equation}
which evaluates to $m_{X} \lesssim m/2$ for $\alpha\theta \sim 50$.
\begin{figure}[h]
\centering
\includegraphics[width=0.95\columnwidth]{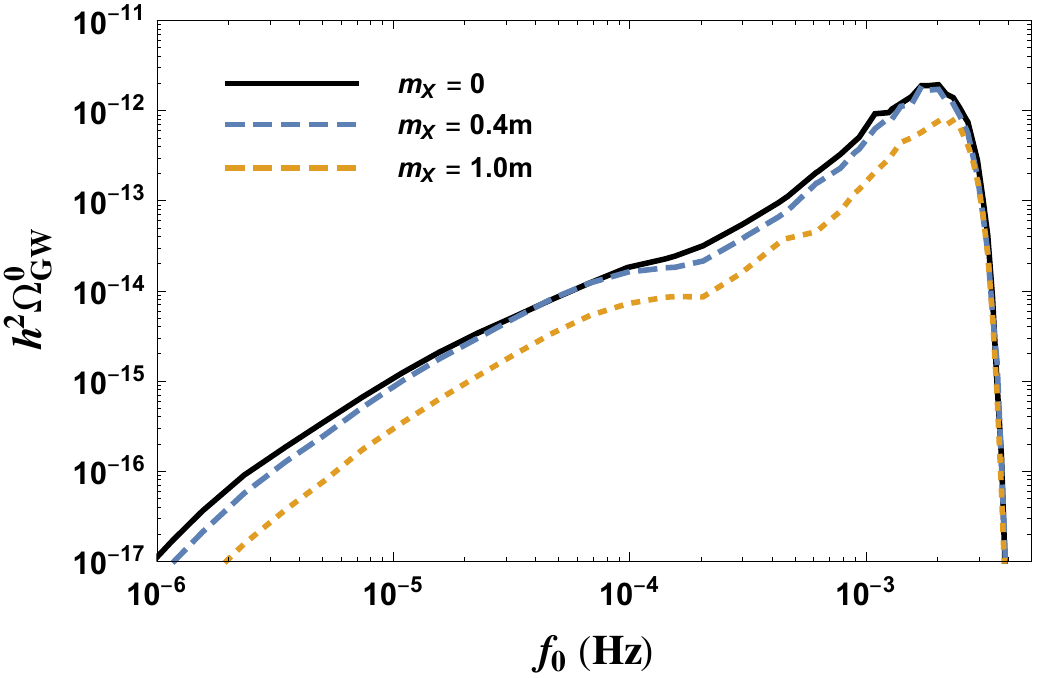}
\caption{Present time GW spectrum for different values of the dark photon mass $m_{X}$  using the improved numerical simulation for the ALP2 benchmark point of Ref.~\cite{Machado:2018nqk}.}
\label{fig:GWspecMx}
\end{figure}
 We show in \cref{fig:GWspecMx} the effect of a massive dark photon on the GW spectrum. One sees that for $m_{X} \lesssim m/2$ that the GW spectrum is largely unaffected. For larger values of the dark photon mass, the tachyonic instability is not as efficient and less energy is transferred from the axion into dark photons and gravitational waves. Thus, the net effect is to damp the GW spectrum.

\section{Relic Abundances and Cosmological Bounds}
\label{app:relics}
The axion starts oscillating when $H_{\text{osc}}=m$, which allows us to infer $T_{\text{osc}}$. At this time the axion energy density is given by
\begin{equation}
 \Omega_{\phi}^{\text{osc}}=\frac{\theta^2 m^2 f^2/2}{3m_{P}^2 H^2}=\frac{1}{6}\left(\frac{\theta f}{m_{P}}\right)^2.
\end{equation}
In the original missalignment mechanism, only redshift $\propto a^{-3}$ has to be taken into account to get the axion relic abundance. Assuming radiation domination and thermal equilibrium for all species ($g_{s}=g_{\rho}$) at the time of oscillation (which holds for $T_{\text{osc}}\gg100$ MeV), the relic abundance without particle production would be
\begin{equation}
 \Omega_{\phi}^0=\Omega_{\phi}^{\text{osc}}\frac{g_{s,\text{eq}}}{g_{\rho,\gamma}}\frac{T_{\text{osc}}}{T_0}\Omega_{\gamma}^{0},
\end{equation}
with $g_{s,\text{eq}}=2+2N_{\text{eff}}(7/8)(4/11)=3.938$, $g_{\rho,\gamma}=2$ and $T_0=2.73$ K. However, the production of dark photons suppresses the axion abundance further. In Fig. 3 of the main text, we indicate using gray lines the amount of suppression particle production must provide (assuming an initial missalignment angle $\theta=1$) in order to prevent over production of dark matter.\\

\subsection{$N_{\rm eff}$ Bound}
An effectively massless dark photon changes the number of effective relativistic degrees of freedom ($N_{\text{eff}}$). At the epoch of recombination, the dark radiation contribution to $N_{\text{eff}}$ is given by 
\begin{equation}
\Delta N_{\rm eff} = \frac{8}{7}\left(\frac{11}{4}\right)^{\frac{4}{3}} \frac{\rho_{X}}{\rho_{\gamma}}\bigg|_{T=T_{\rm rec}} \,,
\end{equation}
where the energy density of SM and dark photons are $\rho_{\gamma}$ and $\rho_{X}$, respectively. Since both species scale as radiation, the fraction ${\rho_{X}}/{\rho_{\gamma}}$ only changes when SM fields become non-relativistic and transmit their entropy to the photon bath. Before transferring its energy to dark radiation, the oscillating axion field redshifts as matter and therefore grows $\propto a$ compared to the SM radiation bath. Following the analysis of Ref.~\cite{Machado:2018nqk}, we assume the majority of the axion energy is transmitted to dark radiation close to the closure of the tachyonic band. For the parameters in question ($\theta\approx1$ and $\alpha\approx50$), tachyonic band closure occurs when the scale factor has grown by an amount
\begin{equation}
 \frac{a_*}{a_{\text{osc}}}=\left(\frac{\theta\alpha}{2}\right)^{2/3} \,,
\end{equation}
which results in the following contribution to $ \Delta N_{\rm eff}$
\begin{equation}
 \Delta N_{\rm eff} = \frac{8}{7}\left(\frac{11}{4}\right)^{\frac{4}{3}}\frac{g_{s,\text{eq}}^{4/3}}{g_{s,*}^{1/3}g_{\rho,\gamma}}\Omega_{\phi}^{\text{osc}}\left(\frac{\theta\alpha}{2}\right)^{2/3}.
\end{equation}
 The Planck 2018 TT,TE,EE,lowE+lensing+BAO dataset constrains $\Delta N_{\rm eff} < 0.3$ at 95\% confidence level~\cite{Aghanim:2018eyx}. This excludes the gray shaded region in Fig. 3 of the main text if the dark photon is effectively massless.
 
\subsection{Dark Photon Dark Matter}
On the other hand, if the dark photon is massive it has the possibility to contribute to dark matter. Note from the dispersion relation for the mode $\tilde k$ that minimizes $\omega^2$ we have 
\begin{align}
 \omega^2(\tilde k)=- \tilde k^{2}  + a_{*}^{2} m_{X}^{2}\,,
\end{align}
so tachyonic production of dark photons requires $\tilde{k} > a_{*} m_{X}$. Therefore we have
\begin{align}
 E > m_{X} \sqrt{1+ (a_{*}/ a)^2} \,,
\end{align}
which is reduced by a factor $\sqrt{2}$ from the time of production ($a=a_*$) to when the momentum becomes negligible due to redshift ($a\gg a_*$). Thus, the dark photon relic abundance is at least suppressed by a factor $1/\sqrt{2}$ compared to the axion abundance in the standard missalignment case. The bound $\tilde{k} > a_{*} m_{X}$ is saturated for $m \sim m_{X}$, which gives the upper bound of the green shaded region in Fig. 3 of the main text.

A dark photon that is light compared to the axion will redshift like radiation for some period of time, resulting in additional suppression of the dark photon relic abundance. In this case, the dark photon mass does not influence the process of particle production (see \cref{app:finiteMass}). Since we assume that particle production takes place when the tachyonic band closes where $\tilde k/a_*=m$, conservation of energy dictates that the number density in dark photons after particle production equals that of the axion before production
\begin{equation}
 n_{\phi}^{\text{osc}}\left(\frac{a_{osc}}{a_{*}}\right)^3=\frac{\theta^2 m f^2}{2}\left(\frac{a_{osc}}{a_{*}}\right)^3=n_{X}^{*}
\end{equation}

Once the dark photon becomes non-relativistic, the relic abundance in dark photons $\rho_X=m_X n_X$ is therefore suppressed by a factor $m_X/m$ as compared to the axion relic abundance in the standard missalignment scenario. For very light dark photons, one has to worry about the bounds on warm dark matter coming from structure formation. We note that the boost factor of a dark photon at matter-radiation equality is
\begin{equation}
 \beta=\frac{\tilde k}{a_{\text{eq}}m_X}=\left(\frac{\theta\alpha}{2}\right)^{2/3}\frac{m}{m_X}\frac{T_{\text{eq}}}{T_{\text{osc}}}\left(\frac{g_{s,\text{eq}}}{g_{s,*}}\right)^{1/3}.
\end{equation}
We infer the bound on the boost factor from studies of sterile neutrino dark matter produced through freeze out, where the average neutrino momentum is given by the temperature. From Refs.~\cite{Narayanan:2000tp, Hansen:2001zv} we find
\begin{equation}
 \beta\lesssim 1.1\times 10^{-4} \,,
\end{equation}
which gives the lower bound on the parameter space where the dark photon can constitute dark matter at $f\approx 3\times 10^{16}$, shaded green in Fig. 3 of the main text.

\section{GW Spectra Fit Parameters}
\label{app:GWfit}
\begin{figure}[ht!]
\centering
\includegraphics[scale=0.6]{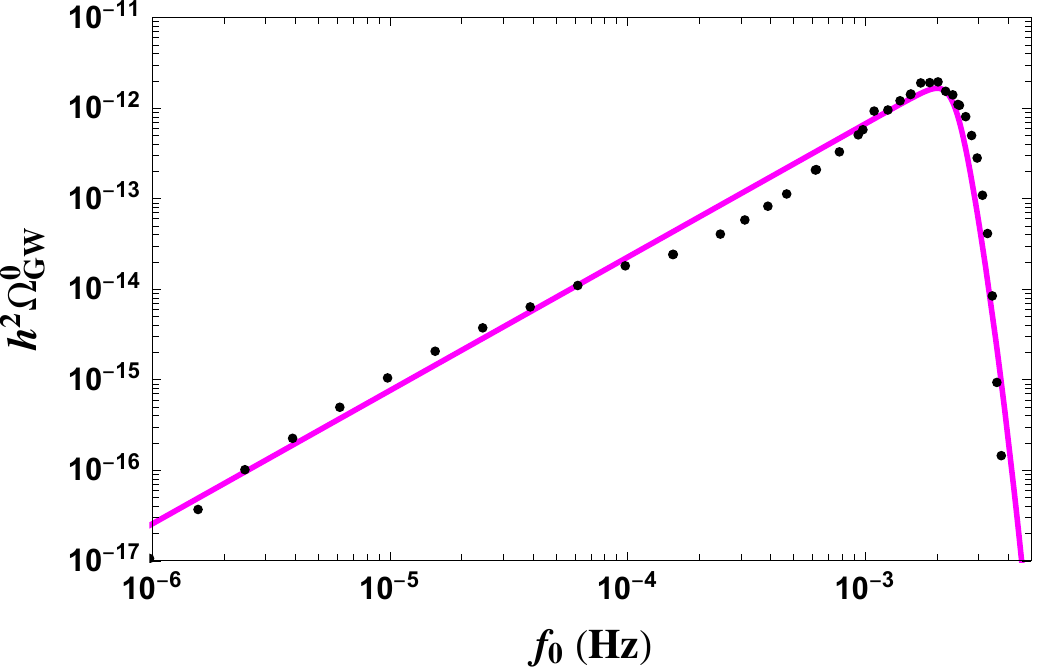} 
\includegraphics[scale=0.6]{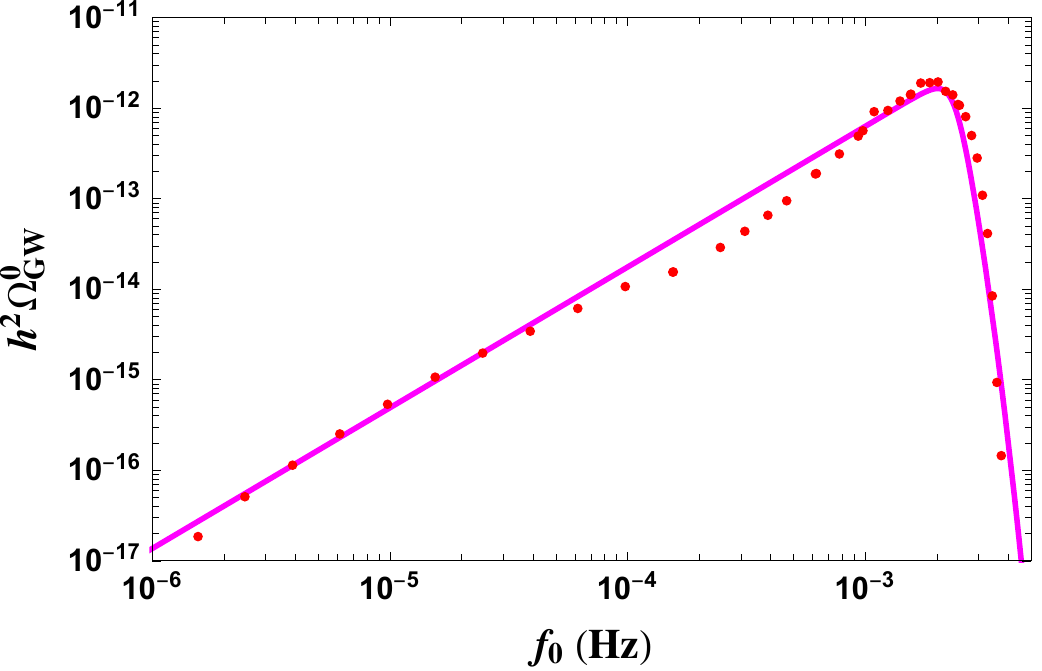}
\includegraphics[scale=0.6]{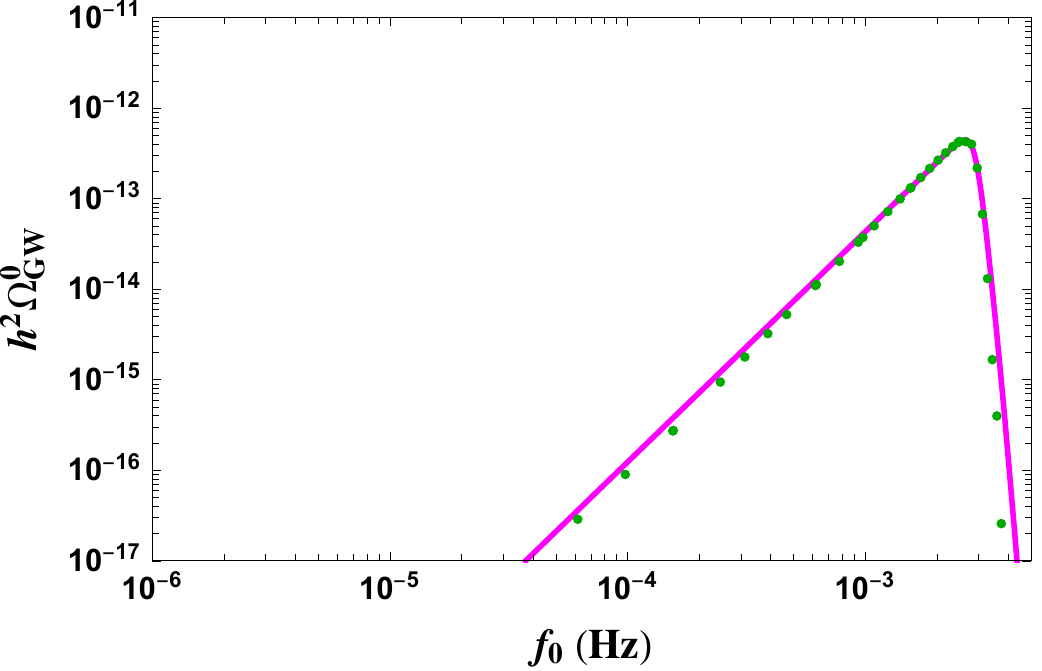}
\caption{Fits to the GW simulation data (magenta) using the template given in  \cref{eq:fit_temp_app}  evaluated using the best fit parameters in \cref{tab:fit_params}. The plots are as follows: Total spectrum (black, top), Dominant chirality spectrum (red, center), and Conservative spectrum (green, bottom).}
\label{fig:fits2data}
\end{figure}

In Section IIIA of the main text, we proposed the following function 
\begin{align}
	\tilde{\Omega}_{\rm GW}(\tilde{f}) = \frac{\mathcal{A}_{s}\left(\tilde{f}/f_{s}\right)^p}{1+\left(\tilde{f}/f_{s}\right)^{p} \exp\left[\gamma (\tilde{f}/f_{s}-1)\right]}\,,
	\label{eq:fit_temp_app}
\end{align}
to fit the gravitational wave spectrum from our simulation. The quantities $\tilde{\Omega}_{\rm GW} \equiv \Omega_{\rm GW}(f)/ \Omega_{\rm GW}(f_{\rm peak})$ and $\tilde{f} \equiv f/f_{\rm peak}$ are the peak amplitude and frequency, which at the time of GW emission are given by
\begin{equation}
f_{\rm peak} \approx (\alpha\theta)^{2/3} m \,, \hspace{2mm} \, \Omega_{\rm GW}(f_{\rm peak})  \approx \left( \frac{f}{M_P} \right)^4 \, \left( \frac{\theta^{2}}{\alpha}\right)^{\frac{4}{3}}  \,,
\end{equation}
and redshifting these quantities to the present time yields
\begin{equation}
f_{\rm peak}^{0} \approx (\alpha\theta)^{2/3} \, T_{0} \left( \frac{g_{s, {\rm eq}}}{g_s}\right)^{1/3} \left( \frac{m}{M_P}\right)^{1/2} \,,
\end{equation}
\begin{equation}
\Omega_{\rm GW}^{0}(f_{\rm peak}^{0})  \approx 1.67\times 10^{-4} g_{s}^{-1/3}\left( \frac{f}{M_P} \right)^4 \, \left( \frac{\theta^{2}}{\alpha}\right)^{\frac{4}{3}} \,,
\end{equation}
where $g_s$ is the number of effective relativistic degrees of freedom associated with the entropy at the time of emission, $g_{s, {\rm eq}} = 3.938$, and $T_{0} = 2.35\times 10^{-13}$ GeV. See Ref.~\cite{Machado:2018nqk} for additional details on the derivation of these formulas. 
We fit the remaining parameters $A_{s}, f_{s}, \gamma, p$ to the spectrum produced by our numerical simulation using a simple least-squares method. We give the results of this fit in \cref{tab:fit_params} for the total GW spectrum, the GW spectrum of the dominant polarization, and the ``conservative" spectrum described in Section III of the main text. To show the goodness of fit, the best fit parameters are plotted using \cref{eq:fit_temp_app} on top of the result from our numerical simulation in \cref{fig:fits2data}.
\begin{table}[h!]
\centering
\resizebox{0.95\columnwidth}{!}{%
\begin{tabular}{|c|c|c|c|c|}
\hline \hline
  &  $A_{s}$  & $f_{s}$  & $\gamma$ & $p$ \\
\hline 
Total spectrum (black) &  6.3 & 2.0 & 12.9 & 1.5 \\ 
Conservative spectrum (green)   &  1.7  & 2.5 & 20.2 & 2.5 \\ 
Left chiral spectrum (red)   &  6.3  & 2.0 & 12.9 & 1.6 \\ 

  \hline \hline
\end{tabular}
}
\caption{Parameter values required to fit our template given in Eq~\ref{eq:fit_temp_app} to the curves shown in \cref{fig:fits2data}.  } 
\label{tab:fit_params}
\end{table}
%
%
	
\end{document}